\documentstyle[aps,prl,amsmath,amssymb]{revtex}
\draft
\begin{document}
 \title{Strong-coupling superconductivity due to soft boson modes 
 in MgB$_2$, cuprates, 
 borocarbides, and some heavy fermions}
\author{D. Lampakis$^1$, A. Tatsi$^1$, E. Liarokapis$^1$,
G. Varelogiannis$^{2}$, P.M. Oppeneer$^3$,
M. Pissas$^4$, 
T. Nishizaki$^5$}
\address{$^1$Department of Applied Mathematics and Physics,
National Technical University of Athens, GR-15780 Athens, Greece\\
$^2$ Institute of Electronic Structure and Laser,
FORTH, P.O. Box 1527, GR-71110
Heraklion, Greece\\
$^3$Institute of Solid State and Materials Research,
P.O. Box 270016, D-01171 Dresden, Germany\\
$^4$Institute for Materials Science, NCSR Demokritos, GR-15310 Athens, Greece\\
$^5$Institute for Materials Science, Tohoku University,
Katahira 2-1-1, Aoba-ku, Sendai 980-8577, Japan} 
\date{\today}
\twocolumn[\hsize\textwidth\columnwidth\hsize\csname@twocolumnfalse\endcsname
\maketitle
\begin{abstract}
We report micro-Raman measurements that 
reveal a low-energy soft mode at 16--19 meV in MgB$_2$
which becomes sharper in the superconducting state.
With a boson spectrum based on this mode
we reproduce, within Eliashberg theory, the $T_c$,
the T-shape of the
NMR relaxation rate as well as unusual
dip-hump features in
available tunneling data.
We compare MgB$_2$  to cuprates,     
borocarbides, and UPd$_2$Al$_3$, where 
we identify soft bosonic modes 
at energies which {\it scale with their $T_c$}, 
suggesting that all these materials are very strong coupling 
superconductors to soft bosonic modes, that are possibly 
precursors of quantum 
instabilities.

\end{abstract}

]
\narrowtext
%

A surprisingly high superconducting (SC)
critical temperature $T_c\approx 40$~K observed 
in a compound as simple as MgB$_2$
\cite{Akimitsu} has stimulated enormous
activity. The simplicity of this system 
gives rise to expectancies that
the origin of the high $T_c$ might be comparatively easy to establish.
Isotope effect measurements \cite{isotope,isotope2} indicate that
phonons participate in the pairing mechanism.
For a material like MgB$_2$,
the phonon spectrum and the electron-phonon coupling are
plausibly
accessible by density-functional theory (DFT) calculations
and a number of such studies
have been reported \cite{DFelphcoupling}.
The main conclusion of all these studies
is that the coupling should essentially
involve the high-energy boron modes 
at around 70 meV 
and the coupling strength
$\lambda$ is
estimated to be in the range $0.4\leq\lambda\leq 0.9$ 
\cite{DFelphcoupling}.
In this range of weak couplings we find some
of the most conventional simple superconductors like Al.

In this Letter,
we report evidence against this
idealized
picture of totally conventional,
weak-coupling superconductivity
in MgB$_2$. 
Our Raman measurements reveal
the existence of a distinct soft
mode in the region 130--150 cm$^{-1}$ (16--19 meV) which
becomes sharper in the
SC state, suggesting its involvement in the
pairing mechanism.
Since all DFT calculations \cite{DFelphcoupling} 
agree that no harmonic
phonon modes
exist at such low energies in MgB$_2$,
we attribute this soft mode 
to the proximity of an unknown 
instability for which DFT methods cannot account. 
We show, using Eliashberg theory
with a boson spectrum
based on this mode, that we reproduce   
the absence of the Hebel-Slichter
peak in the T-shape of the nuclear magnetic resonance (NMR) relaxation rate
in MgB$_2$ as
well as distinct qualitative features
of the available tunneling spectra \cite{Karapetrov,Sharoni,Chen}
(dip-hump structures and anomalous temperature behavior) 
that are characteristic of
{\it exceptionally
strong-coupling superconductivity
in the range
$\lambda\approx 3$}.
We conclude that the observed mode is a signature of the
boson field (in conventional cases harmonic phonons) 
that mediates the pairing.
Since qualitative similar anomalous features in 
the tunneling quasiparticle spectrum
were observed also for high-$T_c$ cuprates \cite{DipCuprates}, 
borocarbides \cite{TunnelBorocarbides},
and UPd$_2$Al$_3$ \cite{Jourdan},
these superconductors are likely to be in the $\lambda \approx 3$
regime as well. 
We identify in all these  materials 
soft modes at energies {\it proportional to 
their respective $T_c$'s}. These soft modes govern the pairing
and are therefore a necessary (but not sufficient)
ingredient for high $T_c$'s.


We have carried out 
micro-Raman measurements 
on high-quality MgB$_2$ powder samples.
The data were obtained with a T64000 Jobin Yvon triple spectrometer
with a liquid-nitrogen cooled CCD.
An Oxford Micro-He cryostat was appropriately modified to
allow the study of microcrystallites at various scattering
geometries. The 488.0 nm Ar$^+$ laser line was focused on
1--2 $\mu$m large crystallites and the power was kept below
$0.03$ mW to avoid heating by the beam.
Typical spectra recorded for selected temperatures are shown in Fig.\,1.  
The main peak at about 600 cm$^{-1}$
was already reported in macro-Raman studies 
\cite{RamanMgB2others,Franck}.
In the low-energy region 130--150 cm$^{-1}$ (16--19 meV)
we observe  
an additional broad peak which becomes sharper
in the SC state.
A structure in the same energy range was
also seen in a macro-Raman study 
\cite{Franck}, but was attributed to the peak in
the quasiparticle density of states 
corresponding to the opening of the
SC gap. However,
in our spectra the feature is clearly visible in the normal state
{\it as well}, therefore we identify it as a soft bosonic mode.
Inelastic neutron
scattering measurements also reported an excitation with a similar
behavior in this energy range (around 17 meV \cite{Sato})
supporting the assignment of the structure to
bosonic excitations. 
Since there is no such soft phonon predicted for MgB$_2$ 
by DFT calculations
\cite{DFelphcoupling}, 
our observed bosonic mode might
result from the proximity to an
instability.
Indeed, a small addition of electrons via Al doping
triggers a subtle structural transition in which the $c$-axis contracts
and $T_c$ drops \cite{Slusky}.
The soft boson mode could be the precursor of this instability.
We argue below that this bosonic mode is likely to be
the principal mediator of the
pairing in MgB$_2$.
We furthermore observe in the SC state a narrow peak
at about 80 cm$^{-1}$ which might be related
to superconductivity. The energy of the latter peak
is too low to be considered a measure
of the SC gap. 

The relevance of our 130 cm$^{-1}$ soft boson for the pairing
mechanism is supported by our Eliashberg calculations
of the NMR relaxation rate $R=T_1^{-1}$ and the tunneling density of states.
NMR has the advantage of being both a local probe and bulk sensitive,
therefore the NMR relaxation data manifest the intrinsic SC character.
We have calculated $R$
under the constraint $T_c\approx 39$~K using an Einstein spectrum
at various energies.                                  
In Eliashberg theory,
the reduced NMR relaxation rate is given by
$$
\frac{R_s}{R_n}=2\int_0^{\infty} d\omega
\biggl( -\frac{\partial f}{\partial \omega}\biggr)
\biggl[\biggl(\Re \biggl\{ \frac{\omega}{    
\sqrt{\omega^2-\Delta^2(\omega,T)} } \biggr\}
\biggr)^2
$$
$$
\qquad\qquad\qquad
+\biggl(\Re \biggl\{ \frac{\Delta(\omega,T)}{     
\sqrt{\omega^2-\Delta^2(\omega,T)} } \biggr\}
\biggr)^2\biggr],
\eqno(1)
$$
where the complex gap function $\Delta(\omega, T)$ results from the   
analytic continuation
to real frequencies 
of the solutions of Eliashberg equations for
imaginary Matsubara frequencies.
We have verified in our calculation that
the NMR relaxation rate is quasi-independent
of the exact spectral shape.
We show in Fig.\,2 the NMR relaxation rate at four
characteristic temperatures 
as a function of the involved boson energy for the two limiting
values of the Coulomb pseudopotential $\mu^*=$0.1 and 
0.2. All experimental points \cite{ExpNMR2}
in the corresponding
temperature range lie within the black rectangle that we
have centered at the energy of our soft boson mode.
Clearly, the absence of the
Hebel-Slichter peak in MgB$_2$ and the behavior of $1/T_1$
near $T_c$
points to frequencies
of the order of our soft mode.
With such low energies involved,
the coupling strength necessary to produce $T_c\approx 40$ K
is sufficiently high $\lambda\geq 2.5$
and the Hebel-Slichter peak is eliminated by
lifetime effects \cite{Rainer}.
We note that recently 
the behavior of the upper critical field in MgB$_2$
\cite{Muller}, and that of the electronic
specific heat \cite{walti01} 
were also proposed as indications of strong-coupling superconductivity.

With an Eliashberg
function based on our boson mode
and adjusted to produce $T_c=39$ K we have calculated
the tunneling density of states.
In the zero temperature
regime we obtain a dip structure 
above the gap (at $\approx 4\Delta(T=0)$)
followed by a hump (see Fig.\,3). As the temperature rises,
the characteristic energies of the structures in
the spectrum (peak-dip-hump) remain almost unchanged, while
the structures broaden gradually. The gap is
filled with states at finite $T$ below $T_c$
contrary to what is expected for            
a conventional second-order phase transition, like in BCS theory
where no coexistence of phases occurs.
Similar unusual behavior of
the quasiparticle density of states was observed so far
in cuprate high-$T_c$ superconductors \cite{DipCuprates},
in UPd$_2$Al$_3$, which is a heavy fermion superconductor
\cite{Jourdan}, and in borocarbides \cite{TunnelBorocarbides}.
In the case of MgB$_2$ a careful examination of
all available tunneling data reveals clear evidence of this
behavior \cite{Karapetrov,Sharoni,Chen}.
For example, by applying a magnetic field, the quasiparticle
density of states in MgB$_2$ (fig. 4 in \cite{Karapetrov})
evolves as already observed
in UPd$_2$Al$_3$ \cite{Jourdan} showing the same anomalous
critical behavior as with temperature in Fig.\,3.
The unusual tunneling spectrum arises
within Eliashberg theory when at
sufficiently strong coupling $\lambda \geq 2.5$ the gap energy becomes
comparable to the energy of the bosons that mediate the
pairing \cite{NatureSato,DipTheory}.
The dip-hump features are not
spectral structures as those reported by McMillan
and Rowell \cite{mcmillan69} but are generic qualitative
features of a very strong-coupling regime at              
the beginning of a cross-over
towards Bose condensation \cite{DipTheory}.

Soft modes as the one we identify in MgB$_2$ 
are shown below to be present in {\it all}
classes of materials that exhibit 
this unusual quasiparticle density of states.
We have made a
detailed Raman study under pressure on
monocrystals of YBa$_2$Cu$_3$O$_{6+x}$ 
compounds. Some of our results are shown in Fig.\,4.
Confirming previous reports \cite{Syassen}, we find in
YBa$_2$Cu$_3$O$_{7-\delta}$ ($T_c = 90$ K)
a soft mode at 320 cm$^{-1}$ {\it
hidden by the B$_{1g}$ mode} (340 cm$^{-1}$) with which it almost
coincides energetically.
By applying pressure, the
energy of the harmonic B$_{1g}$ mode rises as expected, while the
320 cm$^{-1}$ mode remains unchanged \cite{Syassen} 
confirming its soft character. This mode could correspond to the
phonon with abnormal behavior detected at 41 meV by neutron
scattering \cite{Keimer}. The increase in the intensity of
the 320 cm$^{-1}$ soft mode in the SC                     
state may contribute significantly
to the apparent 
softening of the B$_{1g}$ mode with the onset of
superconductivity \cite{Keimer,Friedl}.        

Most surprising is the discovery of a similar soft mode in 
YBa$_2$Cu$_3$O$_{6.5}$ ($T_c=60$ K) at 216 cm$^{-1}$ 
in our micro-Raman measurements under hydrostatic pressures
(cf. Fig.\,4).
In this underdoped compound
the 320 cm$^{-1}$ soft mode has
disappeared and the 216 cm$^{-1}$ 
is the only soft mode we detect.
Surprisingly, the ratio of the energies of the soft modes
almost equals the ratio of the $T_c$'s ($216/320 \approx 60/90$).
In La$_{2-x}$Sr$_x$CuO$_4$ compounds a well-known soft             
mode is identified in the Raman spectra 
\cite{Sugai}.
The frequency of this mode is                    
about 126 cm$^{-1}$ for the $x=0$ compound and it reduces continuously with
doping as does also the temperature where the transition to the tetragonal phase
occurs.
In the case of borocarbides,
the presence of a soft phonon mode was established by neutrons
and widely discussed in the literature
\cite{Stassis}. Finally, even in the heavy-fermion superconductor
UPd$_2$Al$_3$ a soft bosonic mode was observed by neutrons,
which, conversely, was attributed to spin excitations
\cite{NatureSato,Metoki}.

Remarkably,
all these soft 
modes appear at energies that {\it scale with the critical temperature
of the respective materials} (see Fig.\,5). 
The proportionality of these modes to $T_c$ cannot be attributed solely to
gap features because they            
are visible well above $T_c$. The scaling instead suggests
the involvement of these soft
modes in the pairing 
mechanism, and points to {\it a similar, strong-coupling superconductivity
in all these materials}.
The occurrence of strong coupling to soft bosonic modes
in compounds that are so different
raises the question of a common underlying mechanism.
The exceptional strength of the
coupling of electrons to the soft-boson mode
is probably 
due to singularities in the 
quasiparticle susceptibility, that are in turn related to the proximity
of an 
instability. 
The nature of the instability may very well be different in 
different categories of materials.
The proximity of quantum critical points
has already been invoked for
the understanding of superconductivity 
in cuprate \cite{QCPcuprates}
and heavy fermion \cite{QCPhf} materials.
In borocarbides the observed soft modes involve charge degrees
of freedom \cite{Stassis}; 
this is also the case in MgB$_2$ where it
could also be related to the structural instability \cite{Slusky}.
The proximity to an instability produces singularities
not only in the frequency domain responsible for the exceptionally
strong coupling, but also in momentum space, where these may
lead to a SC gap with nodes in some of these compounds.
Anisotropic and even non-s-wave superconductivity in MgB$_2$ or in related
compounds is plausible and compatible with 
the absence of spin degrees of freedom.
In fact, even phonon-mediated
pairing in the 
proximity of a phase separation instability
is dominated by small-q processes, which may lead to anisotropic 
superconductivity 
possibly with unconventional
gap symmetry and other anomalies \cite{smallQ}.

In conclusion, we have detected a soft mode at low energies
($\approx 17$ meV)
in MgB$_2$ which is strongly coupled to the
carriers and 
mediates superconductivity in MgB$_2$.
We show that MgB$_2$ shares an exceptionally 
strong-coupling regime $\lambda \approx 3$ with cuprates, borocarbides
and UPd$_2$Al$_3$. We identify in all these compounds
soft modes at energies which scale with
their $T_c$ and we argue that these are the bosons which govern the
pairing.
Tentatively, the soft modes result in all cases from the proximity
to a (quantum) instability which appears to be a key element for  
obtaining a high critical temperature.

We are grateful to E.N. Economou and K. Prassides
for useful discussions.

\newpage

\begin{figure}[tbp]
\caption{
Typical micro-Raman spectra measured on
MgB$_2$ crystallites. The spectra are corrected for the
Boltzmann thermal factor.}
\end{figure}

\begin{figure}[tbp]
\caption{
The reduced NMR relaxation rate $R_s/R_n$
at four characteristic temperatures near $T_c$ ($T/T_c$=0.95,
0.90, 0.85, and 0.80 from top to bottom
in the low energy side) for $\mu^*$=0.1 (full lines)
and $\mu^*$=0.2 (dashed lines) as a function of the involved
boson energy in MgB$_2$. When $R_s/R_n>1$ 
the Hebel-Slichter peak is present.
The available experimental points for $0.8T_c$ $\leq T\leq$ $0.95 T_c$
lie within the black rectangle centered at the energy of our 
soft boson.}
\end{figure}

\begin{figure}[tbp]
\caption{
The reduced density of states $N_s/N_n$ for various
temperatures
using 
a boson spectrum based on our soft mode whose
amplitude is adjusted to produce
$T_c$=39 K. In the inset we emphasize
the conventional BCS-like T-behavior we obtain
if we use
instead a boson spectrum dominated by
the $E_{2g}$ phonon at 70 meV with the same $T_c$ constraint.
In both cases the temperatures are
T=0.3T$_c$ (full lines), 0.7T$_c$ (dashed lines) 
0.9T$_c$ (long-dashed lines) and 0.95\,T$_c$ (dot-dashed lines).}
\end{figure}

\begin{figure}[tbp]
\caption{
Selected Raman spectra on YBa$_2$Cu$_3$O$_{6+x}$
monocrystals. In the lower panel, the 
320 cm$^{-1}$ soft mode at optimal
doping is evidenced by pressure because
it is insensitive while
the harmonic
B$_{1g}$ phonon at 340 cm$^{-1}$ stiffens.
In the underdoped YBa$_2$Cu$_3$O$_{6.5}$ compound (upper panel)
the 320 cm$^{-1}$ soft mode is absent and 
a new mode at 216 cm$^{-1}$ appears which is
shown to be soft since it remains insensitive to pressure.
In the inset is shown the evolution of the
216 cm$^{-1}$ mode
as a function of the applied pressure.
No other soft mode was detected.}
\end{figure}

\begin{figure}[tbp]
\caption{
The energy of the soft modes identified in 
various materials that show unconventional tunneling
behavior as a function of their respective $T_c$'s.
The linear scaling is evident from the straight fit line
through the origin.}
\end{figure}

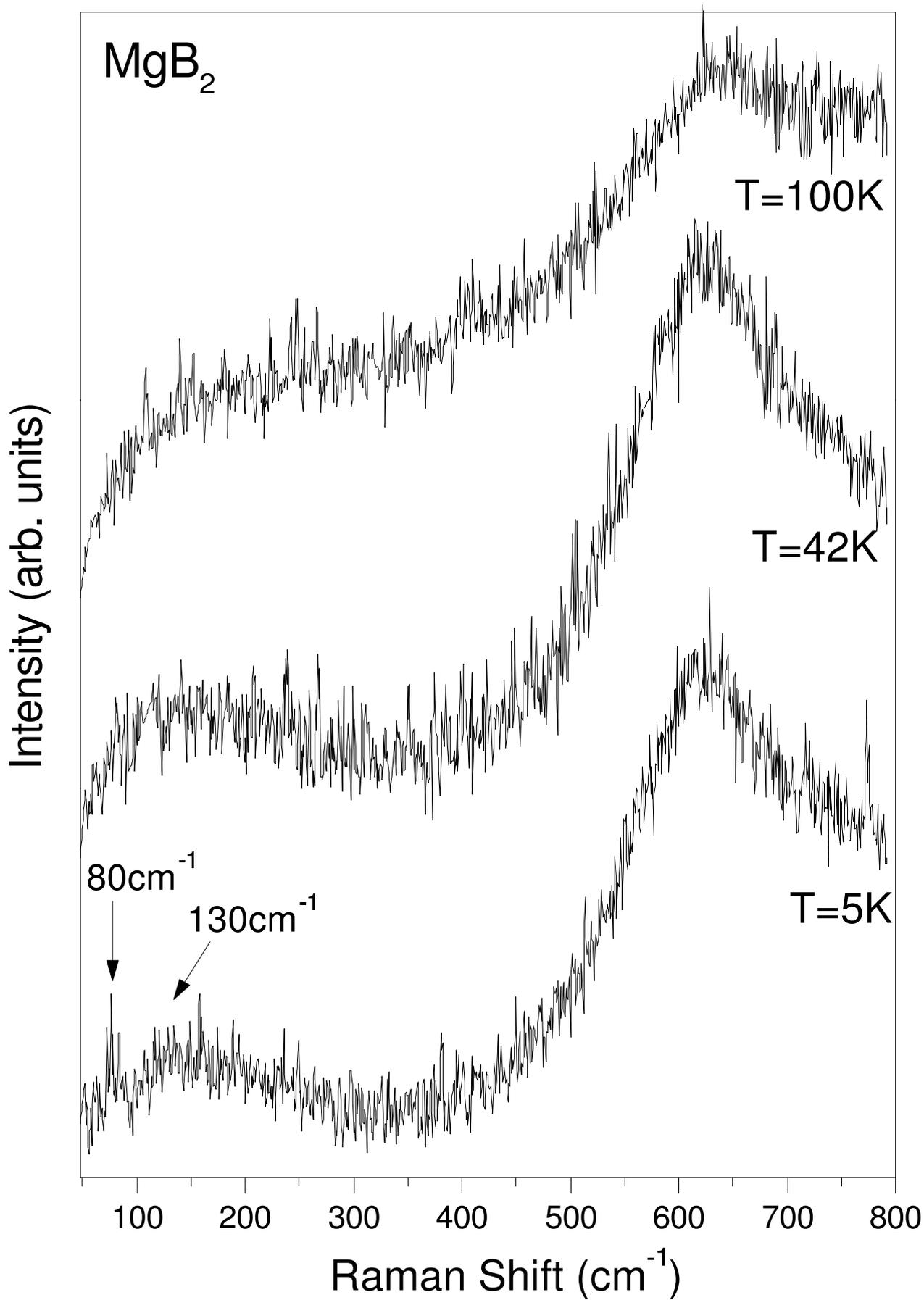
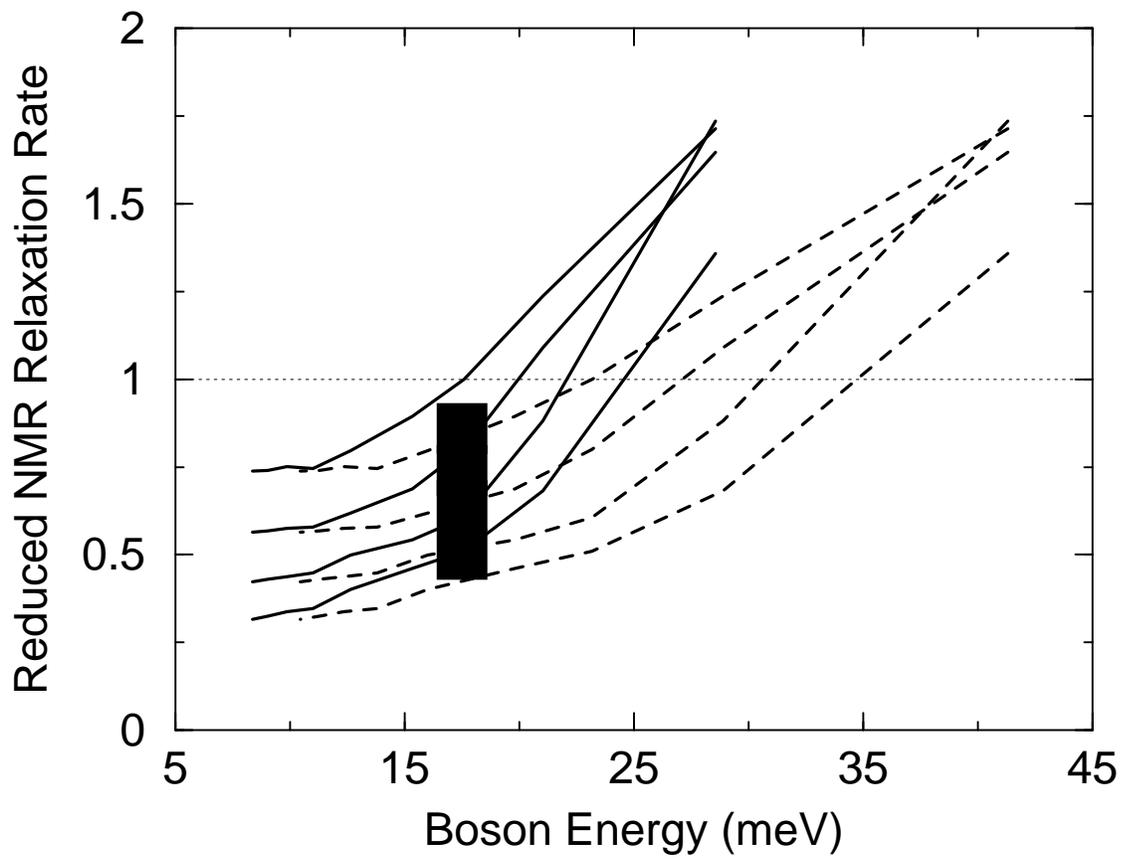
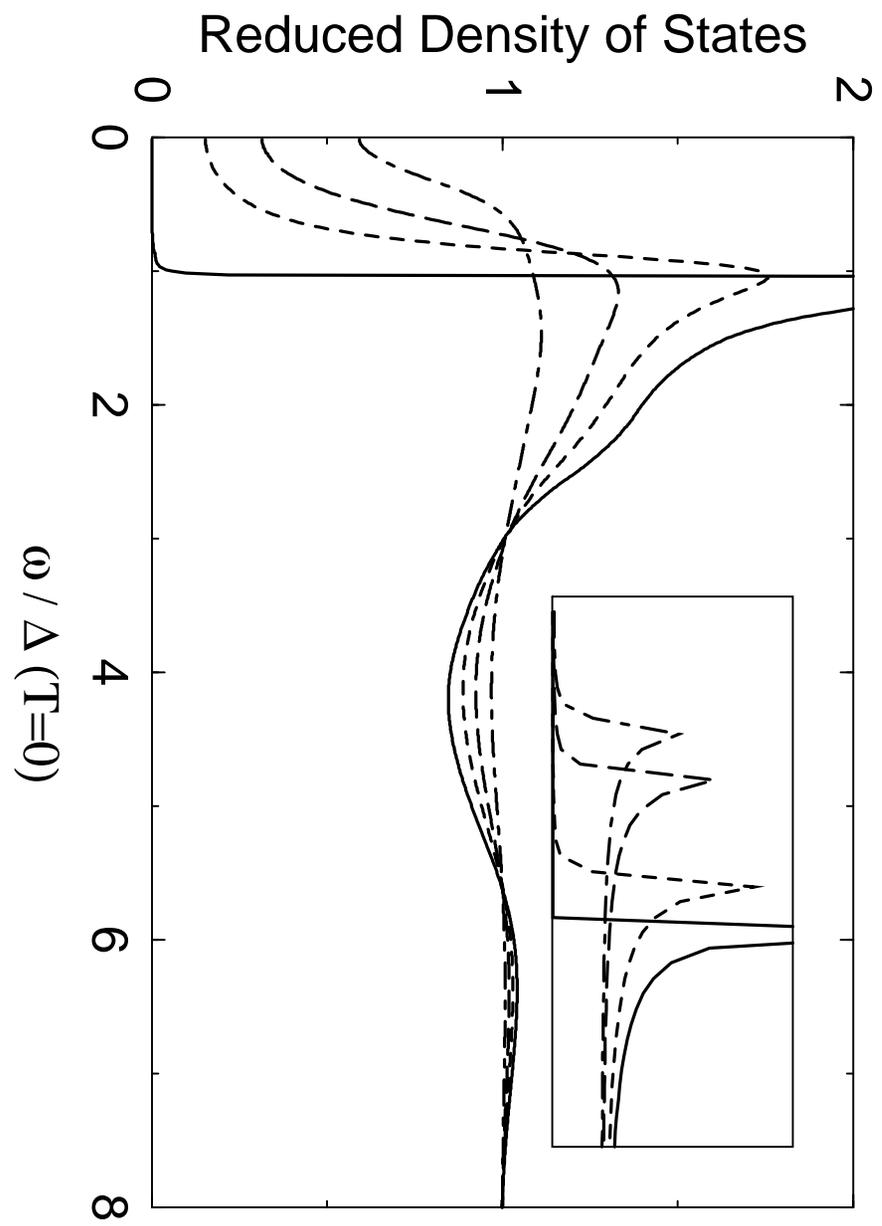
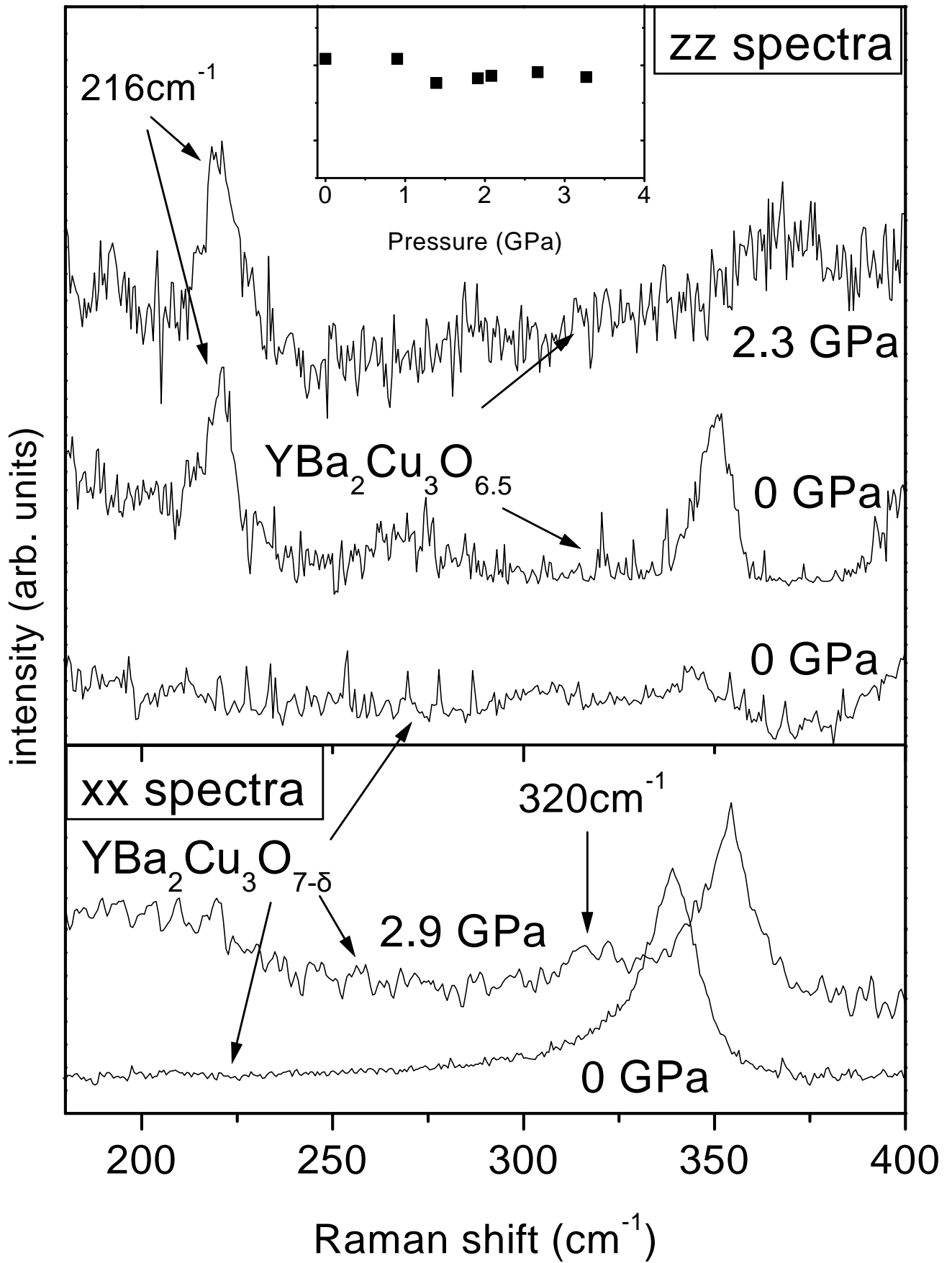
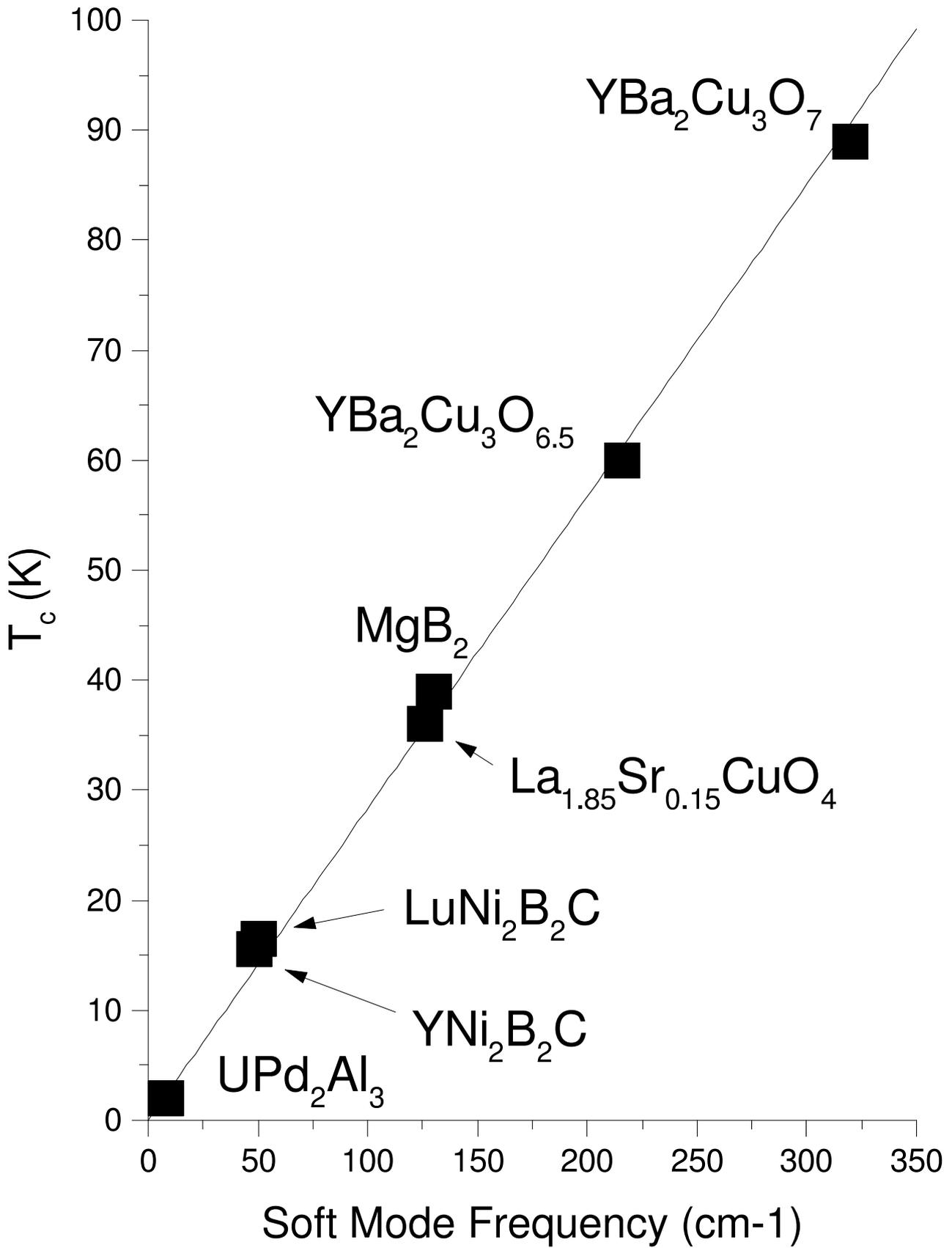
\end{document}